\documentclass[useAMS,usenatbib,twocolumn]{mn2e}

\usepackage{epsfig}
\usepackage[usenames]{color}
\usepackage{graphics}
\usepackage{amssymb}
\usepackage{amsmath}
\usepackage{times}

\title[kHz-QPOs in NS-LMXBs]{Oscillations of relativistic axisymmetric tori and implications for modelling kHz-QPOs in neutron-star X-ray binaries}

\author[Montero, Zanotti]
{P. J. Montero$^{1}$\thanks{E-mail:
montero@mpa-garching.mpg.de}, O. Zanotti$^{2}$ \\
$^{1}$Max-Planck-Institut f{\"u}r Astrophysik,
Karl-Schwarzschild-Strasse 1, D-85748, Garching bei
München, Germany \\
$^{2}$Max-Planck-Institut f{\"u}r Gravitationsphysik, Albert Einstein
Institut, Am M{\"u}hlenberg 1, 14476, Golm, Germany}

\begin{document}

\date{}

\pagerange{\pageref{firstpage}--\pageref{lastpage}} \pubyear{2011}

\maketitle

\newcommand{\be}{\begin{equation}}
\newcommand{\ee}{\end{equation}}
\newcommand{\bdm}{\begin{displaymath}}
\newcommand{\edm}{\end{displaymath}}
\newcommand{\bea}{\begin{eqnarray}}
\newcommand{\eea}{\end{eqnarray}}
\newcommand{\PNM}{P_NP_M}
\newcommand{\halb}{\frac{1}{2}}
\newcommand{\FQi}{\tens{\mathbf{F}}\left(\Qi\right)}
\newcommand{\FQj}{\tens{\mathbf{F}}\left(\Qj\right)}
\newcommand{\FQjj}{\tens{\mathbf{F}}\left(\Qjj\right)}
\newcommand{\nj}{\vec n_j}
\newcommand{\FORCE}{\textnormal{FORCE}}
\newcommand{\GFORCE}{\textnormal{GFORCEN}}
\newcommand{\LF}{\textnormal{LF}'}
\newcommand{\LW}{\textnormal{LW}'}
\newcommand{\WL}{\mathcal{W}_h^-}
\newcommand{\WR}{\mathcal{W}_h^+}
\newcommand{\nur}{\boldsymbol{\nu}^\textbf{r} }
\newcommand{\nuf}{\boldsymbol{\nu}^{\boldsymbol{\phi}} }
\newcommand{\nut}{\boldsymbol{\nu}^{\boldsymbol{\theta}} }
\newcommand{\ar}{\phi_1\rho_1}
\newcommand{\arr}{\phi_2\rho_2}
\newcommand{\ur}{u_1^r}
\newcommand{\uf}{u_1^{\phi}}
\newcommand{\ut}{u_1^{\theta}}
\newcommand{\urr}{u_2^r}
\newcommand{\uff}{u_2^{\phi}}
\newcommand{\utt}{u_2^{\theta}}
\newcommand{\ub}{\textbf{u}_\textbf{1}}
\newcommand{\ubb}{\textbf{u}_\textbf{2}}
\newcommand{\RoeMat}{{\tilde A}_{\Path}^G} 

\newcommand{\pedro}[1]{\textcolor{blue} {\texttt{PM: #1}} }
\newcommand{\olindo}[1]{\textcolor{red}  {\texttt{OZ: #1}} }

\label{firstpage}

\begin{abstract}
We perform a global linear perturbative analysis,
and investigate  the oscillation properties of
relativistic, non-selfgravitating tori 
orbiting  around neutron stars in the slow rotation
limit approximation.   Extending  the  work  done in  
Schwarzschild  and Kerr backgrounds, we consider  the  axisymmetric
oscillations of  vertically integrated tori in the Hartle-Thorne spacetime.
The equilibrium models are constructed by selecting 
a  number of  different non-Keplerian
distributions of specific angular momentum, allowing for
disc sizes $L\sim 0.5 - 600$ gravitational radii.
Our results, obtained after 
solving  a global  eigenvalue problem   to  compute  the  axisymmetric
$p$-modes, indicate that such oscillation modes could account with
most observed lower ($\nu_L$) and upper ($\nu_U$) high frequency quasi-periodic oscillations  for Sco X-1,  and
for some Z sources and Atoll sources with $\nu_L\gtrsim 500$
Hz. However, when $\nu_L\lesssim 500$ Hz, 
$p$-modes oscillations do not account for the linear
relation $\nu_U=A\nu_L+B$, $B\neq0$ between the upper and
the lower high frequency quasi-periodic oscillations that
are observed  in neutron star low-mass X-ray binaries.
\end{abstract}


\section{INTRODUCTION}
\label{introduction}

\label{introduction}

Quasi-periodic oscillations (QPOs) observed in the X-ray
spectra of binary systems are transient phenomena
associated to non-thermal states and state transitions. Those observed
at high frequencies, in the range $\sim 200 \, - 1200
\, {\rm Hz}$~\citep{vanderklis2005}, are referred to as high frequency
QPOs (kHz-QPOs). An important feature of such kHz-QPOs is that they
usually appear in couples consisting of a lower and an
upper kHz-QPO. 
The entire set of kHz-QPOs detected in
black hole binaries (BHBs), i.e. binaries having a black
hole 
as an accretor, contains
only seven sources\footnote{See \citet{Remillard2006}
for a review on X-ray properties of BHBs.}, four of which\footnote{GRS 1915+105,
GRO 1655-40, XTE J1550-564 and H 1743-322.} show both the
upper and the lower kHz-QPO.
On the other hand, there are approximately 
$20$ neutron stars binaries (NSBs), i.e. binaries having a
neutron star as the accretor, that show the  kHz-QPO
phenomenology (see \citet{MillerColeman2010} for a recent
review about QPOs in NSBs), including both Z and Atoll
type\footnote{Based on their timing properties and X-ray spectra, NS-LMXBs are
classified into Z and Atoll sources~\citep{Hasinger89}.}~\citep{Lewin2006}. 
Interestingly, these QPO frequencies correspond to the frequencies of
orbits a few gravitational radii away from a stellar-mass compact
object, which explains why kHz-QPOs have been considered a
promising tool to investigate the gravitational 
forces in the strong field regime. It should also be noted,
however, that in a recent work \citet{Sanna2010} have shown that the
NSB J1701-462 provides an example of a source that,
during the same outburst, presents spectral and timing
characteristic of both Z and Atoll sources. Because the
kHz-QPOs 
detected are remarkably different in the two
spectral states, and because such differences cannot be
attributed to changes in the gravitational field of the
central compact object, the
authors conclude that the coherence and rms amplitude of
the kHz-QPOs cannot be used to deduce the existence of
the innermost stable circular orbit around a neutron star.   

Although it is still not clear whether there is a unique physical
mechanism  responsible for the generation of 
kHz-QPOs in both  BHBs and NSBs, a few
observational evidences have emerged over the years which may
indicate the existence of {\em two distinct} mechanisms. 
\begin{itemize}
\item
First of all, the kHz-QPO peak separation in NSBs
is typically within $20\%$ of the neutron star spin
frequency $\nu_{\rm{spin}}$, or half of
that~\citep{vanderklis2005}. In particular, sources with
$\nu_{\rm{spin}}<400 Hz$ have $\Delta \nu\sim
\nu_{\rm{spin}}$, while sources with
$\nu_{\rm{spin}}>400 Hz$ have $\Delta \nu\sim
\nu_{\rm{spin}}/2$~\citep{Muno2001}. However,
  ~\cite{Strohmayer08} have found that the low-mass NSB 4U0 614+091 does not show a strong connection between the kHz-QPO
frequency difference and the neutron star spin frequency ($\nu_{\rm{spin}}=414$ Hz, while $\Delta \nu\sim 320$ Hz).

\item
Secondly, while in NSBs
the frequencies observed may vary by a factor $2$ in
association to changes of the luminosity, the frequencies
in  BHBs are much more stable and vary at most by $15\%$,
even when the luminosity changes by a factor
$3$~\citep{Remillard2006}. This effect is clearly shown
in Fig. 1 by \citet{Belloni2005} or
in Fig. 3 by \citet{Torok2006},
where the upper frequencies, $\nu_U$, are plotted versus the 
lower frequencies, $\nu_L$; showing that while kHz-QPOs
from a \emph{single} BHB are represented 
essentially as a point,
kHz-QPOs from a \emph{single} NSB are scattered along
straight lines,
the so called
``Bursa lines''~\citep{Abramowicz2007,Torok2006}.

\item
An additional peculiar
property of twin kHz-QPOs in the four BHBs where they have
been simultaneously observed
is that such frequencies
appear in couples obeying the ratio $3:2$ to a high
degree of accuracy (see Fig.1 by \citet{Torok2005b}).
Over the last few years there has been an ongoing debate about the
existence of the same phenomenology among kHz-QPOs in NSBs. 
Observations, in fact, clearly indicate that Bursa lines
$\nu_U=A\nu_L+B$ 
of kHz-QPOs in NSBs are not compatible with a constant
$3:2$ ratio ($A=1.5$, $B=0$), but it remains
controversial whether the peak at $1.5$ in the
distribution of the observed $\nu_U/\nu_L$ ratios is
physical or
not~\citep{Belloni2005,Torok2006,Belloni2007,Torok2008,Boutelier2010}. 

\end{itemize}

Several models have been proposed over the years to
explain the physical mechanism responsible for the
generation of kHz-QPOs in the X-ray spectra of
binary
systems~\citep{Miller1998,Stella1999,Lamb2003,Abramowicz2003,Rezzolla_qpo_03a,Zhang2004,Arras2006}. 
In particular, the model proposed by
\cite{Rezzolla_qpo_03a} for kHz-QPOs 
in BHBs  suggests the existence of an oscillating torus around
the black hole, and identifies the time variability in the X-ray spectra with inertial-acoustic modes ($p$-modes)
of the relativistic thick accretion torus which acts as a resonant
cavity for the $p$-mode axisymmetric oscillations. The frequencies of
the fundamental and the first-overtone modes appear approximately in a
$3:2$ ratio, and within the range of kHz-QPOs depending on the spin of
the black hole and the size of the torus. Moreover, the model by \cite{Rezzolla_qpo_03a} also accounts
for the $M^{-1}$ scaling of the frequency, where $M$ is the mass of
the black hole.

The main properties of the axisymmetric $p$-mode oscillations of 
relativistic, non-selfgravitating tori orbiting  black
holes have been investigated in a series of papers either
using a linear perturbative approach by
\cite{Rezzolla_qpo_03b,Montero2004} (hereafter Paper I
and Paper II, respectively) or through non-linear general relativistic
hydrodynamic simulations of both, non-selfgravitating
tori~\citep{Zanotti03,Zanotti05,Montero07} and also selfgravitating
tori~\citep{Montero08,Montero10}. The
linear perturbative approach, computationally less
demanding, allows for a more detailed investigation of
the parameter space. Overall, those perturbative analysis confirmed
the results of non-linear hydrodynamic simulations which revealed that
the lowest-order eigenfrequencies appear in a sequence of small
integers 2:3:4..., for a wide range of models.
It is also worth noting that some of these modes have
been related in the analysis performed by
\cite{Blaes2006} to surface
gravity waves, at least in the slender torus limit.

However, it remains unclear whether a kHz-QPO model based on
axisymmetric inertial-acoustic oscillations is flexible enough to
explain the rich phenomenology of kHz-QPOs in NSBs, and in
particular to account for the distribution of the
upper and the lower kHz-QPOs along the ``Bursa lines''
observed in NSBs. To address this question we perform a global linear 
perturbative analysis of equilibrium thick discs around neutron stars, whose
external metric is  described in the slow limit approximation. The
set of perturbed relativistic equations is reduced to
the solution of an eigenvalue problem, following the approach
described in Paper I and II. In order to explore exhaustively the deviations
 from the $3:2$ ratio that axisymmetric inertial-acoustic oscillations display, we consider a much wider parameter space than that
studied in Paper I and II, particularly in terms of the rotation
law and the extension of the accretion disc.

 The plan of the paper is as follows: in Section~\ref{GRT} we
introduce the basic assumptions and equations employed in the definition
of our general relativistic, vertically integrated tori. In
Section~\ref{Eqp} we derive the perturbation equations, and in
Section~\ref{Init} we list the properties of the non-selfgravitating
equilibrium models studied. 
In Section~\ref{GRT_ga}, on the other hand, we present
the results of the global analysis, while 
Section~\ref{QPO} is devoted to a discussion of the 
implications for explaining kHz-QPOs in NSBs. 
Finally, Section~\ref{conclusions} contains our conclusions.

In the following we will assume a signature $\{-,+,+,+\}$
for the space-time metric and we will use Greek letters
$\mu,\nu,\lambda,\ldots$ (running from 0 to 3) for
four-dimensional space-time tensor components, while
Latin letters $i,j,k,\ldots$ (running from 1 to 3) will
be employed for three-dimensional spatial tensor
components. We also adopt a geometrized system of
units by setting $c=G=1$.

\section{Equilibrium tori in the Hartle-Thorne spacetime}
\label{GRT}
Equilibrium thick discs around a rotating neutron star 
are constructed assuming that their self-gravity can be neglected and
that the background spacetime takes the form of the Hartle-Thorne
metric \citep{Hartle68}, which describes the metric around a slowly rotating
neutron star. Since we are interested in the region of the spacetime around the
equatorial plane, we use cylindrical
coordinates $(t,\varpi,\phi,z)$, and consider only the zeroth-order terms in the ratio $z/\varpi$ \citep{Wilson1972,Novikov:1973}, where
$\varpi$ is the cylindrical radial coordinate (see Eq.~(1) in
Paper II for this form of the Kerr line element). Then    
the Hartle-Thorne metric in cylindrical coordinates derives from the
Kerr metric, in Boyer-Lindquist coordinates,
by retaining  only those terms which are first order in the ratio
$a/\varpi$, where $a$ is the Kerr parameter, and by replacing the spin
of the black hole with the angular momentum of the
neutron star. In this way we obtain
\bea
\label{a_over_R1}
\Delta&=&\varpi^2-2M\varpi+a^2=\varpi^2(1-\frac{2M}{\varpi}+\frac{a^2}{\varpi^2})
\nonumber \\
\label{a_over_R2}
&&\simeq \varpi^2(1-\frac{2M}{\varpi}) \ , \\
A&=&\varpi^4+\varpi^2 a^2 + 2M\varpi a^2 =\varpi^4(1+\frac{a^2}{\varpi^2}+2\frac{M}{\varpi}\frac{a^2}{\varpi^2})\nonumber \\
&&\simeq \varpi^4 \ , \\
\label{a_over_R3}
\omega&=&\frac{2Ma\varpi}{A}\simeq \frac{2 J}{\varpi^3}
\ ,
\eea
where M is the neutron star mass, and  $J$ is the angular momentum of the neutron
star~\citep{Rezzolla2001}, which we assume to be constant. Then the line element of the Hartle-Thorne metric  becomes
\bea
\label{metric}
ds^2&=&-\left(1-\frac{2M}{\varpi}\right)dt^2 +
\left(1-\frac{2M}{\varpi}\right)^{-1}d\varpi^2 
\nonumber \\
&&- 2\omega \varpi^2 d\phi dt + \varpi^2 d\phi^2 + dz^2
\ .
\eea

In order to construct hydrostatic equilibrium  models of rotating thick discs
we solve
 the continuity equation $\nabla_\alpha(\rho
u^\alpha)=0$ and the conservation of energy-momentum, $\nabla_\alpha
T^{\alpha\beta}=0$, where the symbol $\nabla$ refers to a covariant
derivative with respect to the metric (\ref{metric}). Here,
$T^{\alpha\beta}\equiv(e+p)u^{\alpha}u^{\beta} + p g^{\alpha \beta}$ are
the components of the stress-energy tensor of a perfect fluid, with
$u^\alpha$ being the components of the 4-velocity, $\rho$ the rest-mass
density, $e$ the energy density and $p$ the pressure. 

It is also useful
to introduce an orthonormal tetrad carried by the local stationary
observer and defined by the one-forms with components
\begin{eqnarray}	
{\boldsymbol \omega}^{\hat{t}} = \varpi\sqrt{\Delta/A} {\bf d}t\ ,&\quad
 	&{\boldsymbol \omega}^{\hat{\phi}} = 
  	\sqrt{A}({\bf d}\phi-\omega {\bf d}t)/\varpi\ ,
\nonumber \\ 
{\boldsymbol \omega}^{\hat{z}} = {\bf d}z\ ,&\quad
	&{\boldsymbol \omega}^{\hat{\varpi}} = {\varpi}/{\sqrt{\Delta}}
	{\bf d}\varpi\ .
\label{1-form}
\end{eqnarray}
In this frame, the components of the four velocity of the fluid are
denoted by $u^{\hat{\mu}}$ and the 3-velocity components are defined as
\begin{equation}
v^{\hat{i}}   \equiv \frac{u^{\hat{i}}}
{u^{\hat{t}}} = \frac{\omega^{\hat{i}}_{\alpha} u^{\alpha}}
	{\omega^{\hat{t}}_{\alpha}u^{\alpha}} \ , 
\hspace{1cm} i=\varpi, z, \phi \ .
\end{equation}
We consider a perfect fluid that follows a polytropic equation of
state (EOS) $p = k \rho^{\gamma}$, where $k$ and $\gamma\equiv d\ln
p/d\ln\rho$ are the polytropic constant and the adiabatic index,
respectively. 
Next, following Paper I \& II, we introduce a vertically integrated pressure
\begin{equation}
\label{P}
P(\varpi) \equiv \int_{-H}^{H}p dz,
\end{equation}
and a vertically integrated rest-mass density 
\begin{equation}
\label{Sigma}
\Sigma (\varpi) \equiv \int_{-H}^{H}\rho dz,
\end{equation}
where $H=H(\varpi)$ is the local ``thickness'' of the torus. We further
assume that $P$ and $\Sigma$ obey an ``effective'' polytropic EOS
\begin{equation}
\label{effctv_eos}
P = {\cal K} \Sigma^{\Gamma}\ , 
\end{equation}
so that ${\cal K}$ and $\Gamma\equiv d\ln P/d\ln\Sigma$ play the role of
the polytropic constant and of the adiabatic index, respectively. 

	After the vertical integration, we enforce the conditions of
hydrostatic equilibrium and axisymmetry ({\it i.e.} assume
\hbox{$\partial_t=0=\partial_{\phi}$}) and simplify the equation of
energy-momentum conservation to a Bernoulli-type form~\citep{Kozlowski1978} 
\be
\label{Bernoulli}
\frac{\partial_i P}{E+P} =  -\left(\partial_i \ln u^t -
\frac{\ell}{1-\Omega\ell}\partial_i\Omega\right) \ ,
\ee
where $\ell \equiv -u_\phi/u_t$ is the specific angular
momentum, $\Omega=u^\phi/u^t$ is the angular velocity and 
\begin{equation}
(u^t)^{-2}=-(g_{tt}+2\Omega g_{t\phi} + \Omega^2
g_{\phi\phi}) \ .
\end{equation}
After simple manipulations Eq.~(\ref{Bernoulli}) can be
rewritten as 
\be
\partial_i\ln u^t-
\frac{\ell}{1-\Omega\ell}\partial_i\Omega=\frac{(u^t)^2}{2}\left(\partial_i
g_{tt}   + 2\Omega\partial_i g_{t\phi}+  \Omega^2\partial_i g_{\phi\phi}\right) \ .
\ee
By using the metric terms of~(\ref{metric}) into~(\ref{Bernoulli}), we derive the following force balance equation for
a non-selfgravitating disc in the Hartle-Thorne spacetime
%
%
\bea
\label{vert_eq}
\frac{1}{E+P}\frac{dP}{d\varpi} &=& -\frac{M/\varpi^2-\Omega \varpi (\omega +
  \Omega)}{\left(1-\frac{2M}{\varpi}\right)+\Omega \varpi^2(2\omega-\Omega)},
\eea
where $E$ is the vertically integrated energy density,
defined in complete analogy to \eqref{P} and \eqref{Sigma}.

\begin{table}
\begin{center}
\caption{Main properties of the equilibrium models studied. From left to
right the columns report: the name of the model, 
the type of
specific angular momentum distribution, the constant coefficient
$\ell_{c}$, the power-law index $q$ ({\it cf.}
eq. \ref{powerlaw}) or its range of values,
the minimum radial size for the corresponding sequence of
tori, and the maximum radial size for the sequence of
tori. The angular momentum of the neutron star is set to
be $J=0.1$ in all of the models.}

\label{tab1}
\begin{tabular}{l|c|c|c|c|c|c}
\hline Model &${\rm \ell(\varpi)}$ & $\rm
{\ell_{c}}$ & $q$ & $L_{min}$ & $L_{max}$ \\

\hline
A1    & const.& 3.60 & 0.0 &  0.4 & 1.9 \\
A2    & const.& 3.65 & 0.0 &  0.5 & 5.7 \\
A3    & const.& 3.70 & 0.0 &  0.5 & 10.0 \\
A4    & const.& 3.75 & 0.0 &  0.4 & 16.8  \\
A5    & const.& 3.80 & 0.0 &  0.8 & 29.9 \\
A6    & const.& 3.85 & 0.0 &  0.7 & 71.4 \\
\hline
B1    & power-law & 3.0 & [0.1,  0.15]  & 2.5 & 374.3 \\
B2    & power-law & 3.1 & [0.08, 0.15]  & 2.1 & 348.6 \\
B3    & power-law & 3.2 & [0.07, 0.15]  & 1.9 & 279.6 \\
B4    & power-law & 3.3 & [0.05, 0.15]  & 0.8 & 318.4 \\
B5    & power-law & 3.4 & [0.04, 0.15]  & 2.5 & 363.3 \\
B6    & power-law & 3.5 & [0.02, 0.15]  & 1.7 & 265.3 \\
B7    & power-law & 3.6 & [0.006, 0.15] & 0.4 & 327.1 \\
B8    & power-law & 3.7 & [0.001, 0.15] & 1.9 & 276.7 \\
B9    & power-law & 3.8 & [0.001, 0.15] & 2.6 & 299.1 \\
B10   & power-law & 3.9 & [0.001, 0.15] & 2.7 & 281.6 \\
\hline
C1    & power-law & 2.59 & 0.2  & 3.3 & 160.8 \\
C2    & power-law & 2.15 & 0.3  & 5.3 & 548.8 \\
C4    & power-law & 2.19 & 0.3   & 7.6 & 575.1 \\
C5    & power-law & 2.29 & 0.3   & 1.4 & 550.9\\
C6    & power-law & 2.35 & 0.3   & 1.9 & 545.1\\
C7    & power-law & 2.39 & 0.3   & 0.7 & 156.6\\
C8    & power-law & 1.79 & 0.4   & 51.5 & 593.9\\
\hline

\end{tabular}
\end{center}
\end{table}

\section{Perturbation equations}
\label{Eqp}

	We next perturb the hydrodynamical equations introducing Eulerian
perturbations of the hydrodynamical variables with a harmonic time
dependence of the type
\begin{equation}
\left(\delta V^{\hat{\varpi}}, \delta V^{\hat{\phi}}, \delta Q \right) \sim
	{\rm exp}({-{\rm i}\sigma t})\ ,
\end{equation}
where $\delta Q\equiv\delta P/(E+P)$ and where we have defined the
vertically averaged velocity perturbations respectively as
\begin{equation}
\label{vert_g1}
\delta V^{\hat{\varpi}} \equiv
	\frac{1}{2H}\int_{-H}^{H}\delta v^{\hat{\varpi}}dz \ , \qquad
\delta V^{\hat{\phi}} \equiv
	\frac{1}{2H}\int_{-H}^{H}\delta v^{\hat{\phi}}dz \ . 
\end{equation}
We assume that the Eulerian perturbations in the metric functions can be neglected,
{\it i.e.} $\delta g_{ab}=0$ (Cowling approximation; Cowling,
1941). While this condition does not hold in general, it represents
a very good approximation in the case of non-selfgravitating tori.

To eliminate the imaginary part from the system of equations we
introduce the following quantities
\begin{equation}
\label{var_disp}
\delta U \equiv {\rm i}\delta V^{\hat{\varpi}}\ , \qquad\qquad \delta W
	\equiv \delta V^{\hat{\phi}}\ ,
\end{equation}
and after a bit of straightforward algebra, we derive the
following set of ordinary differential equations
\begin{eqnarray}
\label{euler-rad}
&& \hskip -0.5 cm  
\sigma\frac{\Delta}{\sqrt{A}}\delta U +
	\alpha\frac{\Delta}{\varpi^2} \delta Q'
	+ \left[\frac{\Delta^{3/2}}{A}\left(\frac{A}{\varpi^2}\right)'\Omega 
	\right. -
\nonumber\\
&&\hskip 0.5 cm
	\left. \frac{\Delta^{3/2}}{A}\left(\frac{A\omega}{\varpi^2}\right)'
	+ 2\frac{\Delta^{3/2}}{\varpi^2}(\Omega-\omega)
	\frac{P'}{E+P}\right]\delta W = 0 \ ,
\nonumber\\ 
\end{eqnarray}
\begin{eqnarray}
\label{euler-phi}
&& \hskip -0.5 cm  
\sigma\frac{\varpi^2\sqrt{\Delta}}{A}\delta W + \left\{
	\Omega' + \Omega\ln\left(\frac{A}{\varpi^2}\right)' \right. +
\nonumber\\
&& \hskip -0.25 cm 
	\frac{A}{\varpi^2\Delta}
	\left(\frac{\varpi^2\Delta}{A}\right)'
(\omega - \Omega) -
\nonumber\\
&& \hskip -0.25 cm 
	\left.\frac{A^2\omega'}{\varpi^4\Delta}\Omega^2 - 
	\frac{\varpi^{2}}{A}
	\left(\frac{A\omega}{\varpi^{2}}\right)'\right\}
	\frac{\Delta}{\sqrt{A}}\delta U
	+\frac{A\sigma \alpha}{\Delta
	\varpi^{2}}\left(\omega-\Omega\right)\delta Q = 0  \ ,
\nonumber\\
\end{eqnarray}
\begin{eqnarray}
\label{cont_gr}
&& \hskip -0.5 cm  
\sigma \delta Q + \widetilde{\Gamma}\frac{\Delta}
      	{\sqrt{A}}\delta U'
	+ \Biggl\{\frac{\Delta}{\sqrt{A}}\Biggl[\frac{P'}{E+P}+ 
\nonumber\\ 
&& \hskip 0.0cm 
	\left. 
	\widetilde{\Gamma}\left(\frac{1}{\varpi}-
	\frac{1}{2}\ln\left(\frac{r^2\Delta}{A}+
	\frac{A}{\varpi^2}\Omega(2\omega-\Omega)
	\right)'\right)\right]         \Biggr\} \delta U
\nonumber\\ 
&& \hskip 0.0cm 
 	-\left(\frac{\sigma\sqrt{\Delta}(\omega-\Omega)A\varpi^2}
	{{\varpi^4\Delta}
	+{A^2 \Omega}(2\omega-\Omega)}\right)
	\widetilde{\Gamma}\delta W = 0 \ ,
\nonumber\\ 
\end{eqnarray}
where $\alpha \equiv 1/(u^{t})^{2}$, $\widetilde{\Gamma} \equiv {\Gamma
P}/{E+P}$, and the index $'$ indicates the derivative with respect to $\varpi$.

	Equations (\ref{euler-rad})--(\ref{cont_gr}) are the $\varpi$-
and $\phi$-components of the perturbed relativistic Euler equations and
the perturbed continuity equation, respectively. They can be
solved numerically for the eigenfrequencies and for the eigenfunctions of
$p$-mode oscillations of an oscillating vertically integrated thick disc
in the Hartle-Thorne spacetime. 
In practice, we solve  the system of equations
(\ref{euler-rad})--(\ref{cont_gr}) as an eigenvalue
problem using a ``shooting'' method (Press et al., 1986)
in which, once the appropriate boundary conditions are
provided, two trial solutions are found, starting from
the inner and outer edges of the disc respectively, and these are then matched
at an intermediate point where the Wronskian
of the two solutions is evaluated. This
procedure is iterated until a zero of the Wronskian is found, thus
providing a value for $\sigma$ and a solution for $\delta Q, \delta U$,
and $\delta W$. The numerical method employed here to solve the
eigenvalue problem is the same as that discussed in Paper I and
Paper II, where a more detailed discussion can be found.

\section{Equilibrium models}
\label{Init}

First of all, we need fix the angular momentum of the star which is defined as
$J=I\Omega_\star$, where $\Omega_\star=2\pi/P_\star$ is the
angular velocity of the star. In geometrized units we have
\be
\label{J}
J=\frac{I_{45}}{P_1}\left(\frac{M_\odot}{M}\right)^27.1363\times10^{-4}
\ee
where $I_{45}$ is the moment of inertia of the star in
units of $10^{45}{\rm g} \ {\rm cm^2}$, which we take as $I_{45} = 1.0$, while $P_1$ is the
period of rotation in units of ${\rm \ sec}$. 
For instance, the typical Atoll source 4U 1608-52~\citep{straaten2003}
has a mass $M=1.7 M_\odot$, a spin period of $P=1.61 {\rm ms}$,
and the angular momentum, computed from
Eq.~(\ref{J}), is therefore $J = 0.15$. 
Even assuming a binary system with an
  accreting neutron star that
rotates as fast as the fastest known millisecond pulsar PSR J1748-2446ad, 
namely with P=1.39 msec, and with a canonical mass
$M=1.4 M_\odot$, would yield to $J = 0.26$.

In Sections~\ref{GRT_ga} and ~\ref{QPO} below we report results
obtained after assuming $J=0.1$. 
However, we have also solved the eigenvalue
problem for the case $J=0.2$, without finding
any significant difference in the results, so that our 
conclusions remain unchanged.

Next we define the distribution of the
specific angular momentum $\ell=\ell(\varpi)$ within the
disc. We consider tori with distributions of specific
angular momentum that are constant in space, {\it i.e.} $\ell(\varpi)={\rm
const.}$, and also tori with non-constant  distributions
of the specific angular momentum\footnote{Note that
  \cite{Qian2009} considered an alternative ansatz for the
  distribution of the specific angular momentum.}. 
We note that $\ell(\varpi)={\rm
const.}$ is a useful mathematical case which leads to analytic initial
data, while non-constant  distributions of the specific angular
momentum are a more realistic assumption. In the case of $\ell(\varpi)={\rm
const.}$, the value of the specific angular momentum must satisfy the condition \hbox{$\ell_{\rm ms} < \ell < \ell_{\rm
mb}$}, where $\ell_{\rm ms}$ and $\ell_{\rm mb}$ are the specific angular
momenta of the marginally stable and of the marginally bound orbit in
the Hartle-Thorne spacetime (see e.g. \cite{Abramowicz03b}). On the
other hand, in the case of tori with non-constant distributions of
the specific angular momentum, we consider 
a power-law distribution of the type
\begin{equation}
\label{powerlaw}
{\ell} = \ell_{c}\varpi^{q} \ ,
\end{equation}
where both $\ell_c$ and $q$ are positive constants. The power-law
angular momentum distributions are chosen such that the
position of the cusp is always located between the marginally
bound and the marginally stable orbits. 
The position of the cusp, as
well as the position of the maximum rest-mass density $\varpi_{\rm
max}$ in the torus, are obtained by imposing that the specific angular
momentum at these two points coincides with the Keplerian
value~\citep{Kozlowski1978}. 
The inner edge of the torus
$\varpi_{\rm in}$ is determined by fixing the potential gap
$\Delta W_{in}=W_{\rm in}-W_{\rm cusp}$, defined as
\begin{equation}
\label{potentilagap}
\Delta W_{\rm in}=\ln[(-u_{t})_{\rm in}] - \ln[(-u_{t})_{\rm cusp}]-
        \int_{\ell_{\rm cusp}}^{\ell_{\rm in}}
        \frac{\Omega d\ell}{1-\Omega \ell}\ .
\end{equation}
On the other hand, the outer edge of the torus $\varpi_{\rm out}$
is defined as the position at which
$P=0$ and it is obtained by integration of the hydrostatic balance equation
(\ref{vert_eq}). Then, for a given distribution of specific angular
momentum, sequences of tori having the same $\varpi_{\rm
max}$ but different radial extents can be constructed by varying the potential gap $\Delta W_{in}$. 

\begin{figure}
\hskip 0.5 cm
\includegraphics[angle=0,width=8.cm]{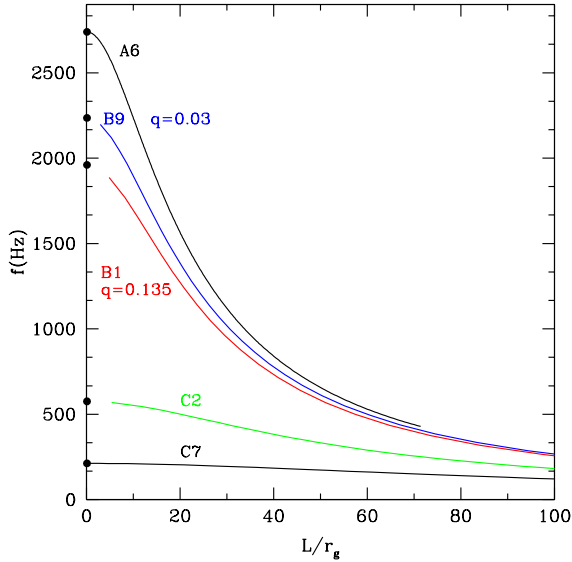}
\caption{Eigenfrequencies for the fundamental mode of axisymmetric $p$
modes for some representative tori with constant and nonconstant
distributions of  specific angular momentum. Each line corresponds to a sequence of tori having the same
$\varpi_{\rm max}$ but different radial extents $L$ and the solid circles
correspond to the values of the Keplerian radial epicyclic frequency
at $\varpi_{\rm max}$ in the Hartle-Thorne metric.}
\label{fig0}
\end{figure}

In order to investigate how the axisymmetric oscillations depend on
the parameters of the discs, we have constructed sequences of models,
having different radial extents and  different distributions of
specific angular momentum. The main properties of the various models
considered are listed in Table~\ref{tab1}. Models of
class A (henceforth models A) are
sequences of equilibrium tori with a constant distribution of specific angular
momentum, while models B and C have a specific angular
momentum increasing outwards according to Eq.~\eqref{powerlaw}.
Unlike models A and C, which correspond to sequences of
discs with different radial sizes for a given pair of values of $\ell_{c}$ and
$q$, models B, refer to different disc sequences which not only have different radial sizes but also 
different values of the power-law index $q$ for each of the
constant coefficients $\ell_{c}$; that is for each set of models from
B1 to B10, we have constructed discs with values
of the power-law index $q$ varying between the minimum and maximum
values listed in the fourth column of Table~\ref{tab1} at intervals
of $\Delta q = 0.005$. The last sequence,  C8,
correspond to discs with a distribution of angular momentum having a power-law
index $q\sim 0.4$, close to the Keplerian value $q_{kep}\equiv
0.5$. Therefore, all these models allow for an extensive
investigation of the parameter space in terms of disc sizes and
distributions of specific angular momentum, varying from constant to
almost Keplerian.

\section{Results of the global analysis}
\label{GRT_ga}

The main properties of axisymmetric oscillations of tori in a
Hartle-Thorne background are analogous to those found for tori in a
Schwarzschild and  Kerr spacetimes (Paper I and Paper
II). Overall, a fundamental mode of oscillation and a
sequence of overtones are found (collectively referred
to as $p$-mode oscillations)
which depend on the  position 
of the rest-mass density maximum, on the radial size of
the disc, on the distribution of angular momentum, while
they  are rather
independent of the equation of state
\citep{Rezzolla_qpo_03b,Montero2004,Zanotti03,Zanotti05,Montero07}. 
These properties can be summarized as follows:

\begin{itemize}
{\item The eigenfrequencies 
 of $p$-mode oscillations increase as the radial size of
  the disc decreases.}
{\item The fundamental-mode tends to
the values of the radial epicyclic frequency at the position of the
rest-mass density maximum as the radial size of the tori tends to zero.}
{\item For any radial extent, the model with the largest
  fundamental-mode eigenfrequency has its rest-mass density
  maximum located at the position at which the epicyclic frequency has
  a maximum.}
{\item The ratio between the frequency of the fundamental-mode ($f$) and its
first overtone ($o_1$) for tori with constant distributions of specific angular
momentum, $\ell(\varpi)={\rm const.}$, appear approximately in a 2:3
ratio. As the size of the tori tends to zero, the ratio $o_1/f$ tends 
to $o_1/f \sim 1.52$. }
{\item The ratio between the frequency of the fundamental-mode ($f$) and its
first overtone ($o_1$) for tori with nonconstant distributions of specific angular
momentum, $\ell(\varpi)= \ell_{c}\varpi^{q}$ can deviate significantly
from $o_1/f \sim 3/2$ for very small discs. As the size of the disc increases the
$o_1/f$ ratio tends to $o_1/f \sim 1.44$.} 
\end{itemize}

\begin{figure}
\hskip 0.5 cm
\includegraphics[angle=0,width=8.cm]{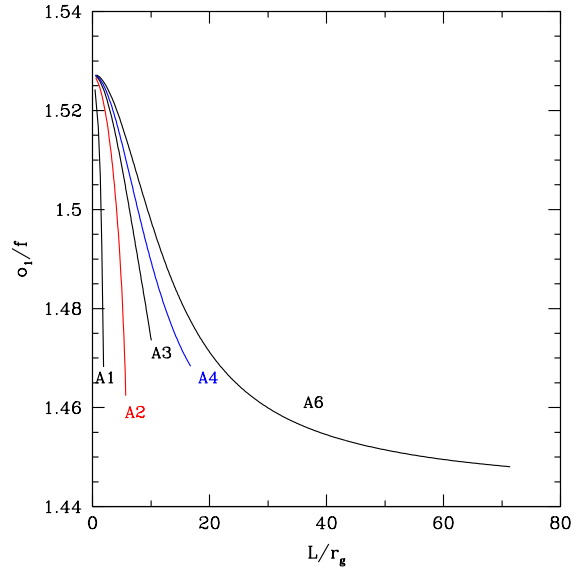}
\caption{Ratio $o_1/f$ as function of the radial
size of the disc for models with a constant distribution of specific
angular momentum, i.e. models A1 to A6.}
\label{fig1}
\end{figure}
\begin{figure}
\hskip 0.5 cm
\includegraphics[angle=0,width=8.cm]{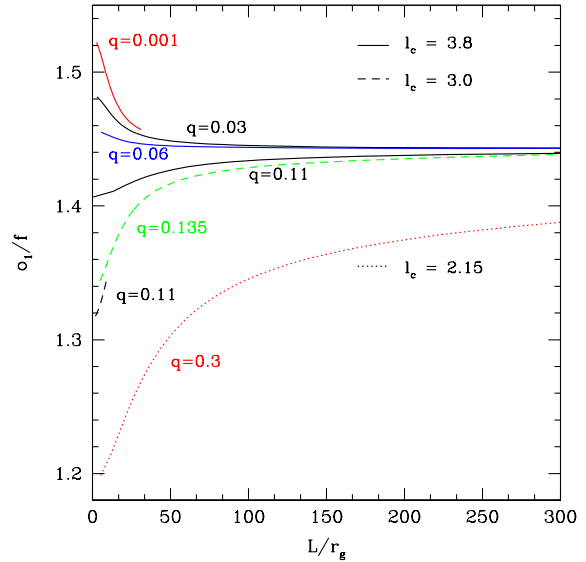}
\caption{Ratio $o_1/f$ as function of the radial
size of the disc for some representative models with a nonconstant distribution of specific
angular momentum, i.e. models B and C.}
\label{fig2}
\end{figure}

\begin{figure*}
\hskip 0.5 cm
\includegraphics[angle=0,width=8.cm]{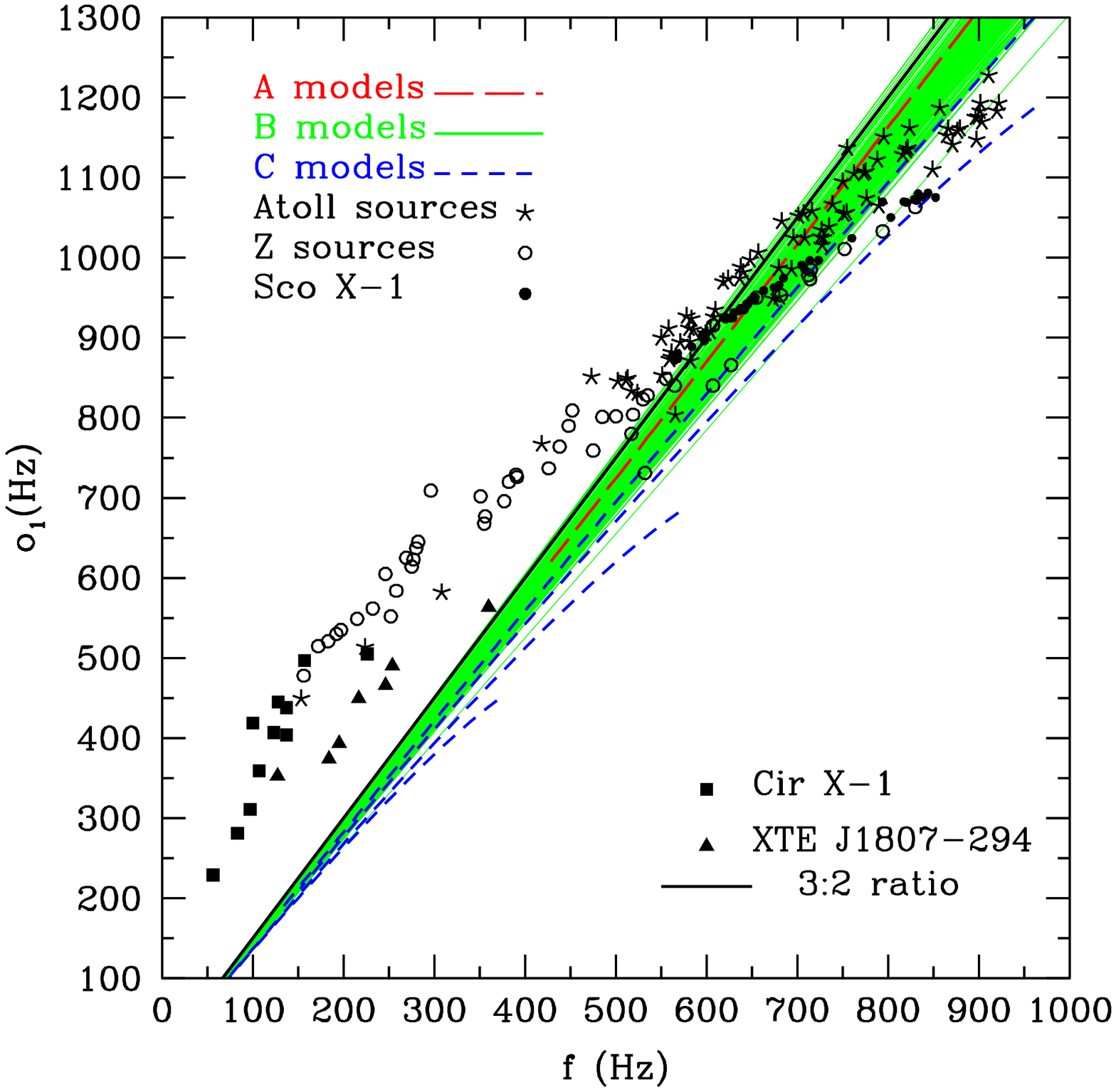}
\includegraphics[angle=0,width=8.cm]{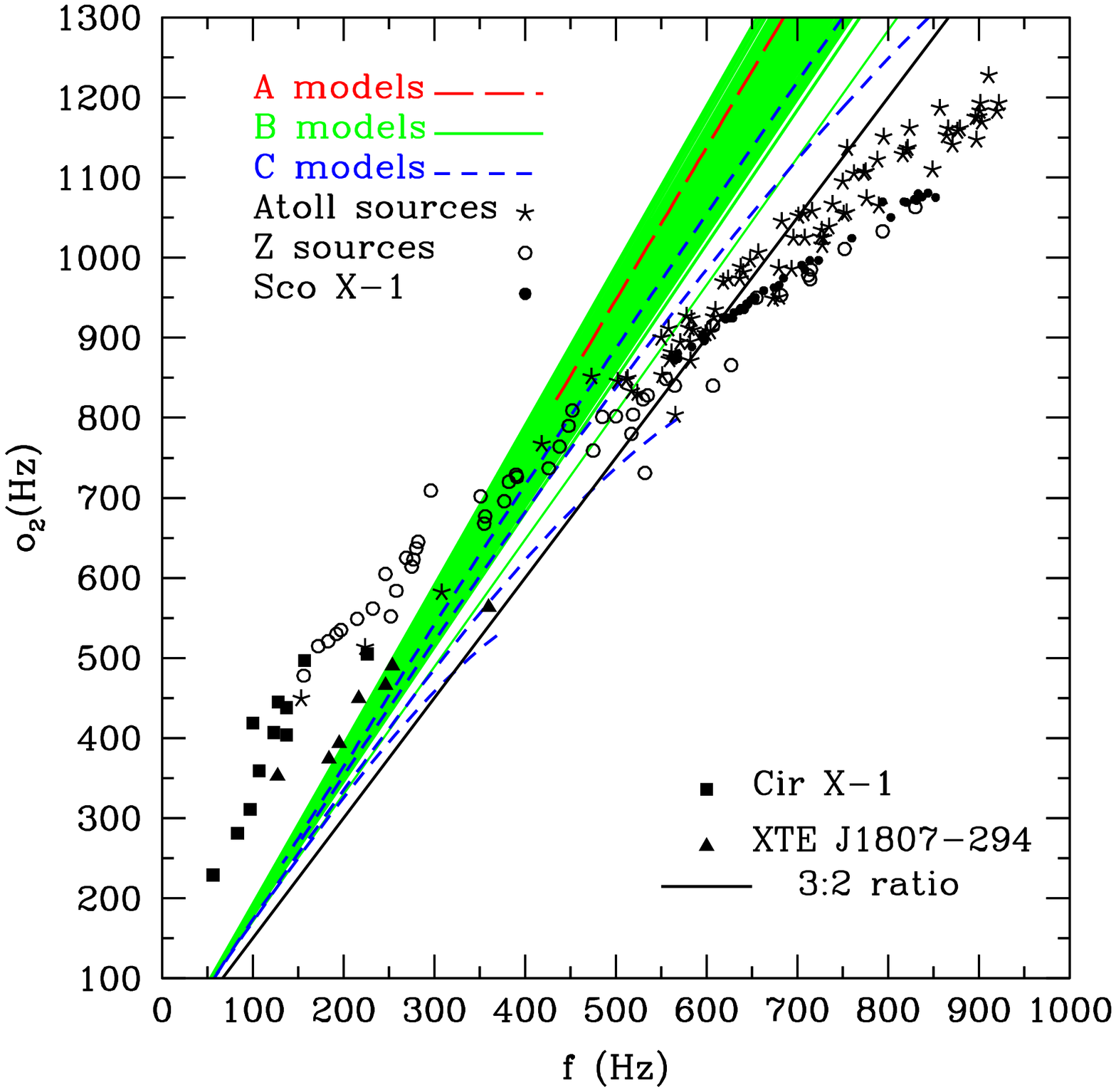}
\caption{Left panel shows a linear plot of the frequency of the upper
kHz QPO versus the frequency of the lower kHz QPO with an asterisk for the Atoll
sources, circles for the Z sources, solid circles for Sco X-1, squares for Cir X-1 and
triangles for XTE J1807-294. We also plot, in the relevant
range of frequencies for kHz QPOs,  the first-overtone versus the
fundamental mode frequencies for all models listed in Table 1. Right
panel shows the second-overtone versus the
fundamental mode frequencies instead.}
\label{fig3}
\end{figure*}

In Figure~\ref{fig0}, we show the eigenfrequencies (in units of Hz and
scaled for a neutron star mass $M=1.6M_\odot$) for the
fundamental mode corresponding to
some representative tori of our sample (i.e., A6, B9 with
$q=0.03$, B1 with $q=0.135$, C2, and C7) and 
reported as a function of the radial
size of the disc expressed in units of the gravitational radii $r_g \equiv GM/c^2$. Each line corresponds to a sequence of tori having the same
$\varpi_{\rm max}$ but different radial extents $L$ and the solid circles
correspond to the values of the Keplerian radial epicyclic frequency
at $\varpi_{\rm max}$ in the Hartle-Thorne metric \citep{Abramowicz03b}.
As expected for modes behaving effectively as sound waves trapped in
the disc, the eigenfrequencies decrease like $L^{-1}$ as the radial extent of the torus
increases. Note also that, as was shown in Paper I and
Paper II for tori orbiting around a Schwarzschild  and a
Kerr black hole, the eigenfrequencies of the fundamental
mode tend to the values of the radial epicyclic frequency
at $\varpi_{\rm max}$ as the radial dimension of the discs tends to
zero. As their size diminishes, 
the role of pressure gradients inside
the disc becomes negligible and
the discs effectively behave as rings of particles
in circular orbits, oscillating with
the epicyclic frequency at the maximum rest-mass density point.

A key feature of the axisymmetric $p$-modes oscillations of tori around
black holes is that the eigenfrequencies of the fundamental mode and the first
overtone appear in an approximately  2:3 harmonic sequence, although
deviations are possible, in particular for the nonconstant specific
angular momentum case. This feature is also present in the case
of sub-Keplerian discs in the Hartle-Thorne spacetime. In
Figure~\ref{fig1} we show the ratio $o_1/f$ as a function of the radial
size of the disc for models with a constant distribution of specific
angular momentum, i.e. models A1 to A6. As the size of the disc
decreases the $o_1/f$ ratio increases and tends to a value of $\sim 1.52$,
independently of the constant distribution of specific angular momentum.  
On the other hand, the behaviour of
the $o_1/f$ ratio, as the disc size decreases, is more complex for nonconstant angular momentum disc. In Figure~\ref{fig2} we plot the ratio $o_1/f$ as function of the radial
size of the disc for some representative models with a
nonconstant distribution of specific 
angular momentum. These models belong to the sequences B and C, and
the values of the constant coefficient $\ell_{c}$ and of the power-law
index $q$ are also shown in Figure~\ref{fig2}. We observe that for
small discs, the $o_1/f$ ratio decreases as the power-law
index $q$ increases for a given value of constant
coefficient $\ell_{c}$, i.e. there exists  a variation
of  about $30\%$ in the $o_1/f$ ratio for small discs. In particular,  $o_1/f$
has an upper limit of $\sim
1.52$ (for discs with an almost constant distribution of specific angular momentum), and a lower limit of $\sim 1.15$ for models with a
power-law index $q$ close to the Keplerian value $q_{kep}=0.5$. On the
other hand, for large-size discs the $o_1/f$
ratio tends to $o_1/f \approx 1.44$.

\section{Implications for kHz-QPOs in Neutron Star Low
  Mass X-Ray Binaries}
\label{QPO}
Based on these properties of the axisymmetric $p$-modes oscillations of
thick discs, \cite{Rezzolla_qpo_03a} proposed a model of
kHz-QPOs in BHBs that explains the observed frequencies
in terms of $p$-modes oscillations of a small accretion
thick disc orbiting close to the 
black hole~\citep{Schnittman06}. This model accounts very well for the 
$M^{-1}$ scaling of the observed frequencies, for the
observed variations in the relative strength of the
peaks, that are interpreted  as due to variations in the
perturbations that the torus is experiencing, and for
the fact that twin kHz-QPOs in the four BHBs show
frequencies obeying the ratio $3:2$ to a high degree of
accuracy. 
As discussed in the Introduction, on the other hand, the
phenomenology of kHz-QPOs in NSBs present peculiar
features that distinguish them from those detected in
BHBs. In particular, the upper and lower kHz-QPOs frequencies
$\nu_U$ and $\nu_L$ can vary by hundred of Hertz along
straight lines $\nu_U=A\nu_L+B$, with $B\neq0$.
For convenience, we have listed in Table 2
the best fit linear parameters obtained by
\cite{Belloni2005,Belloni2007} for the Atoll sources, Z sources, Sco
X-1 and  Cir X-1, highlighting
the linear correlation between $\nu_U$ and $\nu_L$
(errors at 1$\sigma$ significance level).

\begin{table}
\begin{center}
\caption{Best fit linear parameters obtained by Belloni et
  al.(2005,2007) for the Atoll sources, Z sources, Sco X-1 and Cir X-1.}

\label{tab2}
\begin{tabular}{l|c|c}
\hline Source & A & B \\

\hline
Atoll sources    & 0.94 $\pm$ 0.02 & 350 $\pm$ 15 \\
Z sources        & 0.85 $\pm$ 0.01 & 383 $\pm$ 8 \\
Sco X-1          & 0.73 $\pm$ 0.01 & 469 $\pm$ 7 \\
Cir X-1          & 2.34 $\pm$ 0.47 & 104 $\pm$ 58\\
\hline

\end{tabular}
\end{center}
\end{table}

Left and right panels of Figure~~\ref{fig3} show 
the frequency of the upper 
kHz-QPO versus the frequency of the lower kHz-QPO for each
class of available NSBs.
We have indicated  Atoll
sources\footnote{4U 1728-34, 4U 0614+09, 4U 1705-44, KS 1731-260, 4U
  1735-44, 4U 1608-52, 4U1636-53, 4U 1820-30, 4U 1915-05, XTE
  J2123-058.} with an asterisk,
Z sources\footnote{GX 17+2, GX 5-1, GX 340+0, Cyg X-2,
  CX 349-2.} with open circles,
Sco X-1 with solid circles,
Cir X-1 with squares and
XTE J1807-294 with triangles. In addition, for all models for which we
have solved the eigenvalue problem and that
are listed in Table 1,  we have plotted,
in the relevant range of frequencies for kHz-QPOs,  
the first-overtone versus the
fundamental mode frequencies (left panel) and
the second-overtone frequency against
the fundamental mode frequency (right panel).
As it is shown in both panels of 
Figure~\ref{fig3}, the portion of
the plot covered by the observational data intersect only
marginally with the values obtained from the eigenmode analysis.
In particular, we show on the left panel, that there are discs (mostly B models) with fundamental mode frequency
$f > 500$ Hz and a first-overtone frequency which can be in agreement
with most observed $\nu_L$ and $\nu_U$ kHz-QPO frequencies for Sco X-1,  and
for some Z sources and Atoll sources. This range of fundamental mode
frequencies ($f > 500$ Hz) indicates that the
corresponding suitable models would be small in
size (L smaller than $\approx 50{\rm r}_g$), and would have a
nonconstant distribution of specific angular momentum (B models).

On the contrary, we do not  find  models lying above the
line with constant slope $3:2$, which would be
needed to explain the observed twin QPOs with
$\nu_L\lesssim 500$ Hz. As shown in
Figure~\ref{fig0}, the fundamental mode frequency
decreases as the size of the disc increases or as the
distribution of specific angular momentum approaches the
Keplerian profile. The $o_1/f$ ratio tends to $o_1/f \sim 1.44$  as the
size of the disc increases or to smaller values for small discs with
(see Figure~\ref{fig2}). This reflects in the tendency, that models
show, to concentrate towards the $3:2$ ratio line as the
fundamental mode frequency tends to zero (left panel of
Figure~\ref{fig3}).

The possibility that the observed $\nu_L$ and $\nu_U$ kHz-QPO frequencies
correspond to the fundamental frequency and to the second overtone of
an oscillating torus, encounters similar difficulties (see right panel of
Figure~\ref{fig3}). Although the area
covered by the computed $p$-modes oscillations  match some of the
observed kHz-QPO frequencies, particularly those of Z sources with 
$\nu_L$ in the range $150-500$ Hz, the observations of most of the Atoll sources,
of Cir X-1 and of several Z sources remain unexplained. 

Overall, the properties of axisymmetric $p$-mode
oscillations of vertically integrated thick discs are such that, in a
plot $\nu_U$ versus $\nu_L$, the first-overtone and the fundamental mode frequency follow a straight
line for which $A$ may depart  from an exact $3:2$ ratio by $30\%$
(in the case of discs with a nonconstant distribution of
the specific angular momentum), but for which $B \approx 0$.

\section{Conclusions}
\label{conclusions}

We have performed a detailed analysis of the oscillation
properties of a thick disc (torus) around a slowly
rotating neutron star. Our approach extends previous
investigations by \cite{Rezzolla_qpo_03b} and
\cite{Montero2004} by considering a much wider parameter
space and by solving the linear perturbative eigenvalue
problem in the Hartle-Thorne metric. In particular,
the rotation law of the torus spans the whole range
between a constant distribution of the specific angular
momentum and an almost Keplerian rotation. 
We have computed a fundamental mode of oscillations and
a sequence of overtones which can in principle be all
excited depending on the perturbation acted upon the
torus. 

We showed there are discs (B and C models) with fundamental mode frequency
$f \gtrsim 500$ Hz and a first-overtone frequency which can be in agreement
with most observed $\nu_L$ and $\nu_U$ kHz-QPO frequencies for Sco X-1,  and
for some Z sources and Atoll sources.  

However, when these results are used for explaining kHz-QPOs in neutron star low mass
X-ray binaries with twin QPOs with
$\nu_L\lesssim 500$ Hz, a major difficulty arises. In fact,
unlike kHz-QPOs in black hole binaries, the upper and
the lower kHz-QPOs in
neutron star binaries obey a 
linear relation $\nu_U=A\nu_L+B$, with $A$ significantly
different from $1.5$ (e.g. $A\approx 0.94$ for Atoll sources, $A\approx
0.85$ for Z sources and $A\approx 0.73$ for Sco X-1) and $B\neq0$. On the contrary, 
the  computed axisymmetric $p$-modes, either in the ratio
$o_1/f$ or $o_2/f$ follow a straight line with $0.8 \lesssim A
\lesssim 1.5 $ and with $B\approx 0$ for $o_1/f$, and with $ A \gtrsim
1.5 $  and $B\approx 0$ for $o_2/f$.

Therefore, with the assumptions adopted in this paper,  
axisymmetric $p$-modes oscillations of a thick disc around a neutron
star do not provide an explanation for the observed twin QPOs in
neutron star X-ray binaries with $\nu_L\lesssim 500$ Hz.
Nevertheless, additional physics should be taken into
account to have a better understanding of the differences between the Z
and Atoll sources. For instance, the thickening of the disc due
to radiation pressure, the interaction
of the accreting thick disc with the magnetosphere of the
neutron star, the presence of a magnetic
field~\citep{Balbus91}, or non-axisymmetric 
instabilities~\citep{Papaloizou84,Kiuchi11} may play an important
role. 

It is known that the luminosities of the Z sources are typically close to the Eddington
luminosity ($L\sim L_{\rm EDD}$), while the luminosities of the Atoll sources appear
in a lower and broader range ($L\sim 0.001-0.5 L_{\rm EDD}$). Moreover, \cite{Hasinger89} first suggested
 that the differences among the two type of sources
may be due to differences in the mass
accretion rates, i.e. low and high accretion rates show the source as
an Atoll or a Z one. \cite{Lin09} found that the source XTE J1701-462 evolved from super-Eddington luminosity to
quiescence, displaying an evolution with features of Cyg-like Z,
Sco-like Z and Atoll sources, supporting the idea that
changes in the accretion rate are responsible for this secular
evolution. In addition, \cite{Lin09} pointed out that the Atoll stage may be
characterized by a constant inner disc radius, while the Z stage may
have a luminosity dependent location of the inner disc. Similar
results have been recently obtained  by \cite{Ding11}, who focused on the
interaction between the NS magnetosphere and the
radiation-pressure-dominated accretion disc. As the disc is thickened
by radiation pressure, the disc gas pressure reduces and the
magnetosphere expands, thus making the inner disc radius
increase.

 Overall, there is a growing evidence that
the different source stages are
due to different disc structures, and that the hot regions of the flow
where the QPO oscillations are generated could be qualitatively
different. Interestingly, the oscillation properties of
relativistic, non-selfgravitating vertically-integrated equilibrium
tori orbiting NSs agree better with the data for the
Atoll (low accretion rate) stage and the Sco-like stage than for 
the Z (high accretion rate) stage. This seems to indicate that
both the  NS magnetosphere interaction and the thickening
of the disc due to radiation pressure may play a crucial
role, and will be considered in our future research for a
more realistic interpretation of kHz-QPO from neutron
star binaries.

\section*{Acknowledgments}
We would like to thank E. M{\"u}ller and L. Rezzolla for their
comments and careful reading of this manuscript. We also thank
T. Maccarone for useful discussions. It is also a pleasure to thank Shin Yoshida for his contributions to the
numerical code used to solve the eigenvalue problem, and Tomaso Belloni for providing
the observational data used in Fig.~\ref{fig3}.
We also thank an anonymous referee for very helpful comments.
PM acknowledges support from the Deutsche Forschungsgesellschaft
(DFG)  through its Transregional Center ``Gravitational Wave Astronomy'' SFB/TR 7.

\bibliographystyle{mn2e}
\bibliography{aeireferences}

\bsp

\label{lastpage}

\end{document}